\documentstyle[11pt]{article}

\newcommand{\om}{\omega}

\newcommand{\pa}{\partial}
\newcommand{\be}{\begin{equation}}

\newcommand{\ee}{\end{equation}}
\newcommand{\bea}{\begin{eqnarray}}
\newcommand{\eea}{\end{eqnarray}}
\begin{document}
\begin{center}
{\bf{\Large An Analytic Study of the Phase Structure of Lattice QCD with
Wilson Fermions at Infinitely Strong Coupling.}}\\[1cm]
{ A. Galli}\\[0.3cm]
{\em Paul Scherrer Institute, CH-5232 Villigen PSI, Switzerland}\\[0.3cm]
{\today}\\[1cm]
{{\bf Abstract}}
\end{center}
The phase structure of lattice QCD with two flavors and
Wilson fermions is studied analytically.
At $\beta=0$ we obtain rigorous lower and upper bounds for the
critical hopping parameter $k_c(0)$ from a convergent
hopping parameter expansion
to infinite order. The result supports the value $k_c(0)=\frac{1}{4}$ observed
in
Monte Carlo simulations.
\newpage
\section{Introduction}
The outstanding properties of QCD are asymptotic freedom and
confinement. Although a proof that QCD is confining does not exist, there are
strong indications that QCD has this property and we assume it to be true in
the
following. The coexistence of asymptotic freedom and confinement means that
the effective quark-gluon
and gluon-gluon coupling changes continuously from weak coupling at short
distance to strong coupling at long distance.
At long distance quarks and gluons are confined inside color singlet hadrons.
All present known strong interaction phenomena support the confining picture.
In addition hadrons built from the two lightest quarks exhibit an approximate
flavour symmetry which is spontaneously broken.
Due to the confinement, a perturbative description of these
phenomena is not possible. The description of the long distance strong forces
requires non-perturbative methods. Lattice QCD formulates the theory
non-perturbatively and allows us to test confinement and to determine
numerically properties of low energy hadrons. \\

Some useful qualitative informations in lattice QCD can be obtained in the
infinitely strong bare gauge coupling limit $\beta=0$. At infinite gauge
coupling confinement and spontaneous breakdown of chiral flavour symmetry are
manifest. The spontaneous chiral symmetry breakdown and the emergence of
massless Goldstone bosons explain the lightness of the pion.
The small but non vanishing
mass of the pion is due to the non exactness of the flavour symmetry due to the
small but finite light quark masses.\\

In the past many Monte Carlo
simulations of lattice QCD with Wilson fermions
has shown the existence of a critical line $k_c(\beta)$
in the $k-\beta$ plane, where the mass of the Goldstone boson
vanishes. For any given coupling $\beta$ there exists a critical hopping
parameter
$k_c(\beta)$. The continuum limit is approached along this line which ends at
$k_c(\infty)=\frac{1}{8}$.
In the strong coupling regime, at $\beta=0$,
Monte Carlo simulations indicate that $k_c(0)=\frac{1}{4}$.
Some strong coupling
expansion analysis give also some good argument
to support this feature \cite{seiler,strong}.\\

Up to now there is not a rigorous proof that $k_c(0)=\frac{1}{4}$ for Wilson
fermions.
On the other hand, at infinite gauge coupling an important simplifiction of
the system takes place which makes feasible
a convergent hopping parameter expansion analysis to {\em infinite} order.
In fact, at $\beta=0$ all gauge variables on different links are
independent and can be integrated out  explicitly, leaving us with a pure
fermion problem which is however still not trivial.\\

In this work we analyse rigorously the behaviour of the Goldstone boson
spectrum at $\beta=0$ with the method of the hopping parameter
expansion.
Our analysis is based on a random
walk representation of the meson propagators \cite{Fro}. In particular this
allows
us a rigorous estimation of the mass of the pseudoscalar isotriplet and of
$k_c(0)$.
A full hopping parameter expansion to infinite order of perturbation is a very
difficult problem because there are no known general methods to identify
all diagrammatic contributions to the expansion for a given order of
the perturbation expansion.
However, there is
an easy way to determine an upper and a lower bound to $k_c(0)$ to infinite
order of the perturbation expansion
by neglecting and overcounting, respectively, a class of
diagrammatic contributions of the expansion which are very difficult to be
identified. This simplifiction leads to an upper and a lower estimation of
the mass of the Goldstone boson (the pseudoscalar isotriplet)
expressed simply by geometric series of the hopping parameter $k$. These
bounds allow us to determine bounds on the critical hopping parameter $k_c(0)$
from the condition that this mass has to vanish.

\section{The Wilson lattice action}
There is a SU(3) color matrix U(b) in the fundamental representation defined
on each oriented lattice bond b. Our convention is that
\be
U(-b)=U^\dagger(b).
\ee
An oriented path $\om$ on the lattice is a set of bonds
\be
\om=b_1\cup b_2\cup\dots\cup b_n
\ee
such that the endpoint of $b_i$ is the start point of $b_{i+1}$ for
$1\leq i\leq n-1$. We can associate a SU(3) color matrix with $\om$ by defining
the path ordered product
\be
U(\om)=U(b_1)U(b_2)\dots U(b_n)
\ee
The spin matrices are defined in terms of $\gamma$-matrices by
\be
\Gamma(b)=\left\{\begin{array}{cc}
\Gamma^\mu=r+\gamma^\mu&\mbox{if b in $+\mu$ direction}\\
\bar\Gamma^\mu=r-\gamma^\mu&\mbox{if b in $-\mu$ direction}
\end{array}\right.
\ee
The $\gamma$-matrices are hermitian 4x4 matrices, satisfying
$\{\gamma^\mu,\gamma^\nu\}=2\delta^{\mu\nu}$.\\
The Wilson action is then defined on a lattice $\Lambda$ by
\be
S=\beta\sum_{p\subset\Lambda}TrU(\pa p)-k\sum_{b=\langle xy\rangle}\bar\psi(x)
\Gamma(b)U(b)\psi(y)-\sum_{x\in\Lambda}
\bar\psi(x)\psi(x)
\ee
where $p$ represents a plaquette.
The quark fields are represented by the anti-commuting variables $\psi(x)$ and
$\bar\psi(x)$ which transform under the $3$ and $\bar 3$ representation of
color. We consider only two flavors.\\

\section{The meson propagator}
In order to evaluate masses of mesons we have to
consider the propagator of these
states. The masses are obtained by studying the long distance behaviour of
these
propagators. A propagator of a meson is given by
\be
\langle M(x)^\dagger M(y)\rangle=\frac{1}{Z}\int [d\bar\psi][d\psi][dU]
M(x)^\dagger M(y)e^{-S(U,\bar\psi,\psi)}
\ee
where $Z$ is the partition functional and
$M(x)=\bar\psi(x)M\psi(x)$ represents a meson operator ($M$ is some
matrix in spin, flavour and color space).
The hopping parameter expansion is obtained by breaking the action $S$ into two
parts: the unperturbated part $S_0$ and the perturbation $S_I$.
\bea
S_0&=&-\sum_{x\in\Lambda}\bar\psi(x)\psi(x)\nonumber\\
S_I&=&-k\sum_{b=\langle xy\rangle}\bar\psi(x)\Gamma(b)U(b)\psi(x).
\eea
Here the plaquette term is not present because we are at $\beta=0$.
Expanding the exponential in eq. (6) in term of
the perturbation and integrating out the gauge degrees of
freedom $U$ we obtain an expression for the the meson propagator
\bea
\langle M(x)^\dagger M(y)\rangle &=&\nonumber\\
&=& \sum_{\omega^+:x\mapsto y}
C_M\times Tr\left(M^\dagger\Gamma(\omega^+)M\Gamma(\omega^-)\right)\times
(-1)^L
k^{|\omega^+\cup \omega^-|}\equiv \nonumber\\
&\equiv& \sum_{\omega^+:x\mapsto y}C_M\times
A(\omega^+\cup\omega^-)
\eea
where $C_M$ is some overall constant
amplitude which does not affect the meson mass,
$\omega^+$ is a path on the lattice,
$\omega^-=-\omega^+$ is a path in the opposite direction of $\omega^+$
and $L$ is the number of closed loops formed by the path
$\omega^+\cup \omega^-$. Here $\Gamma(\omega)$ and $U(\omega)$ represent the
ordered products of $\Gamma(b)$ and $U(b)$ on the bonds $b\in\omega$,
respectively.
We do not give a proof of (8) since it is standard \cite{Fro,creutz}.

\section{Meson masses from the hopping parameter expansion}
In order to compute the meson masses we consider the static propagator
\be
G^M(t)=\sum_{\vec x}\langle M((0,\vec x))^\dagger M((t,\vec x))\rangle.
\ee
The lowest order diagram representing a static propagator is a double fermion
path $\omega_0$ in the time direction (Fig.1). Since there is a non-zero
probability for a
transition from the lowest order state to a more complicate state, the full
propagator is given by a sequence of exited states connected by lowest
order
states (Fig.2). Each double path $\omega\neq \omega_0$ can be viewed as a
space-time process contributing to the excitation. These excitations
renormalize
the mass of the static propagator.\\
The unrenormalized mass of a meson can be found from the properties of the
lowest order diagram of the static propagator which is given by
\be
G_0^M(t)=C_M\times \exp(-m_0t)=C_M\times(4k^2)^t.
\ee
for $r=1$ in (4).
This tells us that the unrenormalized mass is $m_0=-log(4k^2)$.
Having computed the unrenormalized static propagator,
we will now compute the renormalization
produced by intermediate exited states. Fig. 3 shows us such an excitation,
with initial point $w$ and final point $z$, where the static
unrenormalized propagator arrives and
departs. For fixed $w$ and $z$, we will sum over all intermediate exited states
to
obtain a total weight for the event. We denote this weight by
$ D^{ M}(w,z)$. The
full propagator takes the form \cite{Fro}
\bea
\langle M(x)^\dagger M(y)\rangle&=&\sum_{w_i,z_i(i=1\dots
n)}G_0^{{ M}}(x,w_1) D^{ M}(w_1,z_1)G_0^{{ M}}(z_1,w_2)
D^{ M}(w_2,z_2)\times\nonumber\\
& &\times\dots \times  D^{ M}(w_n,z_n)G_0^{{ M}}(z_n,y)
\eea
where the points $w_i$ and $z_i$ are required to lie between x and y and to be
ordered so that each $z_i$ is "later" than $w_i$, which is itself "later" than
$z_{i-1}$. This last equation represents the picture in Fig. 2.\\
In order to find the renormalized mass of the propagator, we consider
the full static propagator
\be
G^{{ M}}(t)=\left.\sum_{x,y}\langle M(x)^\dagger M(y)\rangle\right
|_{y_0-x_0=t}
\ee
where $x_0$ and $y_0$ represent the time component of the lattice points $x$
and $y$, respectively.
Similarly we define\footnote{Notice that
$(G_0^{ M}(t))^{-1}=\frac{1}{C^{ M}} e^{m_0t}=\frac{1}{C^{ M}} (4k^2)^{-t}$}
\be
\tilde D^{ M}(t)=(G_0^{ M}(t))^{-1}\left.\sum_{x,y}
D^{ M}(x,y)\right|_{y_0-x_0=t}
\ee
Here we have normalized the excitation relative to a static unrenormalized
propagator over the time interval $t$.
{}From this definition and eq. (11) we obtain
\bea
G^{{M}}(t)&=&e^{-m_0t}
\sum_{s_i,t_i(i=1\dots n)}\tilde D^{ M}(t_1-s_1)\dots \tilde D^{ M}(t_n-s_n)=
\nonumber\\
&\simeq&e^{-m_0t}\,e^{p_{{ M}}t}
\eea
where $t_i$ and $s_i$ are the corresponding time coordinates of $z_i$ and
$w_i$. The exponential representation of (14) is up to irrelevant
boundary effect exact.
So the renormalized mass $m_{{ M}}$ is
\be
m_{{M}}=m_0-p_{{ M}}.
\ee
The leading order of $p_{ M}$ can just be written in the form
\be
p_{{M}}=
\sum_{z=(t'\geq 0,\vec z)}(G_0^{ M}(t'))^{-1}D^{ M}(0,z)=
\sum_{z=(t'\geq 0,\vec z)}(4k^2)^{-t'}D^{ M}(0,z)
\ee
To proof this last equation we have to insert it into the last term of (14) and
to expand the exponential function containing $p_{ M}$, then we have to
compare the result with the first term of (14). The excitation term
$D^{M}(w,z)$ is defined starting from eq. (8) in the following way:
\be
D^{M}(w,z)=\sum_{\om:w\mapsto z\mapsto w}\,\,
\frac{A(\om)}{C_{M}}
\ee
where the sum is over all closed paths $\om=\om^+\cup\om^-$ from $w$ to $z$
and return.
$A(\om)$ denotes the amplitude of an intermediate
excitation diagram and $C_M$ is the overall constant of eq. (8) and (10).
\section{Goldstone boson mass}
The critical hopping parameter $k_c(0)$ can be evaluated from the condition
that the mass of the Goldstone boson (the pseudoscalar isotriplet)
vanishes. The mass of the Goldstone boson
is expressed in term of a hopping parameter expansion serie by eq. (15-17).
The Goldstone boson is characterised by the pseudoscalar isotriplet bound state
matrix $M=M_{Gb}\equiv\gamma_5T^I$ for $I=1,2,3$,
where $T^I$ are the isospin SU(2) generators. \\
To evaluate the correction to the unrenormalized mass $m_0$ due to the
excitation $p_M$ we have to identify all closed paths $\om$ from the point $0$
to
the point $z$ and to sum over all points $z=(t\geq 0,\vec z)$ with positive
time in eq. (16). An excitation is a closed path. At $\beta=0$ we do not have
plaquettes in the expansion because the plaquette expectation value vanishes,
therefore an excitation is a closed path consisting
of a double line $\om=\om^+\cup\om^-$
with $\om^-=-\om^+$ composed by a set of connected bonds
\be
\om=\om^+\cup\om^-=\left[\bigcup_{b_i\in\om^+}b_i\right]\cup\left[
\bigcup_{b_i\in\om^+}(-b_i)\right].
\ee
An excitation has to be irreducible, which means that cutting a double bond
$b\cup-b$ of
the path $\om$ in the {\em time} direction, the resulting two paths have some
{\em time} overlap. Reducible excitations can be split into several time
separated
excitations, therefore to avoid overcounting we have to consider only
irreducible
excitations. \\
Using the following two properties for $r=1$ of the gamma matrices
$\Gamma(b)\Gamma(b)=2\Gamma(b)$ and
$\Gamma(b)\Gamma(-b)=0$
the reader can easily convince himself that the amplitude $A(\om)$ of an
excitation of the Goldstone boson can be expressed by a product of terms
depending on the order of the
bonds which compose the path $\om$
\be
A(\om=\om^+\cup\om^-)=C_{M_{Gb}}\times A_0
\times \prod_{b_i\subset \om^+}F(b_{i},b_{i+1})k^2
\ee
where $A_0$ and $C_{M_{Gb}}$ are some constants and
\be
F(b_i,b_{i+1})=\left\{\begin{array}{cl}
0&\mbox{ if $b_{i+1}=-b_i$}\\
2&\mbox{ if $b_{i+1}\perp b_i$}\\
4&\mbox{ if $b_{i+1}\parallel b_i$}
\end{array}\right.
\ee
The hopping parameters is squared in (19) because the product is only on the
half path $\om^+$ and all the bonds of $\om^-=-\om^+$ are paired with bonds in
$\om^+$.\\
There are two classes of irreducible paths $\om$. The first contains paths with
all
bonds $b_i$ perpendicular to the time direction. These paths are trivially
irreducible, because they have no bonds in the time direction. It is very easy
to
identify and count them. The second class contains irreducible paths which have
some
bond parallel to the time direction (Fig. 4).
These paths are very difficult to be
counted because they are irreducible and therefore they can not be decomposed
into a product of spatial excitations separated in time
and time-like double bonds.\\
Because the function $F$ in (20) is positive
we can however undercount and overcount the irreducible paths
contributing to the sum in
(17), by neglecting all paths of the second class and, respectively,
considering all possible paths (reducible and irreducible) with
bonds in all directions\footnote{We recall that considering in the sum (17)
a reducible paths has the effect of overcounting some irreducible paths.}.
We obtain in this way an  underestimation and an overestimation of
$p_{M_{Gb}}$:
\bea
p_{M_{Gb}}^{under}&=& A_0
\times \sum_{S'\subset S}
\prod_{b_i\in S'}C(b_{i},b_{i+1})F(b_{i},b_{i+1})k^2\\
p_{M_{Gb}}^{over}&=& A_0
\times \sum_{S'\subset S}\sum_{T'\subset T}
\prod_{b_i\in S'\cup T'}C(b_{i},b_{i+1})F(b_{i},b_{i+1})k^2
\eea
where the set $S$ represents all bonds perpendicular to the time direction and
the set $T$ all bonds parallel to the time direction.
The term $(4k^2)^{-t'}$ in the sum (16) is not present in (21) because there is
no time extension in the sum in (21). In (22) it is implicitly present in the
overestimated form $(k^2)^{-t'}$.
The mapping
$C(b_{i},b_{i+1})$ controls if the sequence of bonds forms a connected path
\be
C(b_i,b_{i+1})=\left\{\begin{array}{cl}
0&\mbox{ if $b_{i+1}$ and $b_i$ are sequentially disconnected}\\
1&\mbox{ if $b_{i+1}$ and $b_i$ are sequentially connected}\\
\end{array}\right.
\ee
We define two bonds $b_i$ and $b_{i+1}$ to be sequentially connected if the end
point of $b_i$ is the start point of $b_{i+1}$.
It is easy to see that the set of {\em all} possible paths which contribute
non-trivially to eq. (21-22) forms a 3-dimensional and, respectively, a
4-dimensional tree. The
number of branches of these trees counts the number of different paths.
For a n-dimensional tree each bond can be chosen perpendicular or
parallel or antiparallel to the previous one. For all these possibilities the
function $F$ is known and given by (20). Because of the particular form of $F$
we obtain for (21) and (22) two geometrical series\footnote{We consider,
for example,
eq. (22). For a given bond $b_i$ we can choose
$b_{i+1}$ perpendicular ($2(d-1)$ possibilities) or parallel ($1$ possibility)
or antiparallel (1 possibility) to $b_i$. Using the value of the function $F$
we obtain for these two bonds: $2\times 2(d-1)+4\times 1+0\times 1=4d$.
Eq. (21) is analogous to eq. (22).}:
\bea
p_{M_{Gb}}^{under}&=& A_0
\sum_{n=1}^{\infty}(4(d-1)k^2)^n\\
p_{M_{Gb}}^{over}&=& A_0
\sum_{n=1}^{\infty}(4dk^2)^n
\eea
where d=4 is the dimension of the space-time lattice and $A_0=\frac{1}{4}$ is
evaluated explicitly from eq. (8). \\
The final result is an over and upper bound on the Goldstone boson mass.
{}From eq. (15) we obtain
\be
-\log(4k^2)-\frac{dk^2}{1-4dk^2}\leq
m_{M_{Gb}}(k,\beta=0)\leq -\log(4k^2)-
\frac{(d-1)k^2}{1-4(d-1)k^2}
\ee
{}From the condition $m_{M_{Gb}}=0$ we obtain two bounds on $k_c(\beta=0)$
\be
0.231<k_c(\beta=0)<0.264.
\ee
This rigorous result supports the observed value in lattice Monte Carlo
simulations.\\
The same method can be used for finding an upper and a
lower bound on the masses of the $\rho$ and $\omega$ mesons.
The $\rho$ and $\omega$ mesons are characterised by the matrices
$M_{\rho}=\gamma^jT^I\,\,\,(j=1,2,3)$
and $M_{\omega}=\gamma^j\,\,\,(j=1,2,3)$, respectively.
One has to find the corresponding functions
$F$ in eq. (20).
For other mesons the method can not be applied because the function $F$
vanishes.

\vspace{0.5cm}
{\Large {\bf Figure Caption}}
\begin{enumerate}
\item A static unrenormalized propagator.
\item An excitation of the static unrenormalized propagator.
\item A close up of an intermediate excitation.
\item A typical irreducible excitation with some bond in the time direction.
\end{enumerate}
\end{document}